\documentclass[pre,twocolumn,superscriptaddress,preprintnumbers,amsmath,amssymb]{revtex4-1}
\usepackage{epsfig}
\usepackage{color}
\usepackage{amsmath}
\usepackage{amsfonts}
\usepackage{graphicx}
\usepackage{fancybox}
\usepackage{subfigure}
\usepackage{notes2bib}
\usepackage{fontenc}
\usepackage{amssymb}
\usepackage{caption}
\usepackage[version=3]{mhchem}

\usepackage{subfiles}
\begin{document}


\title{Water Density Fluctuations Relevant to Hydrophobic Hydration are Unaltered by Attractions}


\author{Richard C. Remsing}
\author{Amish J. Patel}
\email{amish.patel@seas.upenn.edu}
\affiliation{Department of Chemical and Biomolecular Engineering,
University of Pennsylvania, Philadelphia, PA 19104}

\date{\today}

\begin{abstract}
An understanding of density fluctuations in bulk water has made significant contributions to our understanding of the hydration and interactions of idealized, purely repulsive hydrophobic solutes.
To similarly inform the hydration of realistic hydrophobic solutes that have dispersive interactions with water, here we characterize water density fluctuations in the presence of attractive fields that correspond to solute-water attractions. 
We find that when the attractive field acts only in the solute hydration shell, but not in the solute core, it does not significantly alter water density fluctuations in the solute core region.
We further find that for a wide range of solute sizes and attraction strengths, the free energetics of turning on the attractive fields in bulk water are accurately captured by linear response theory.
Our results also suggest strategies for more efficiently estimating hydration free energies of realistic solutes in bulk water and at interfaces.

\end{abstract}

\maketitle

\raggedbottom


\section{Introduction}
Hydrophobic effects and solvent-mediated phenomena in general, are important in a wide variety of contexts~\cite{Chandler:Nature:2005,ball08,berne_rev09,Jamadagni:ARCB:2011}
ranging from protein folding~\cite{dill_rev02,Levy:2006aa,Thirumalai:2010aa} and aggregation~\cite{Dobson:2003,shea08,Thirumalai:2012},
to colloidal assembly~\cite{Whitesides:2002aa,Rabani:2003,Morrone:2012} and detergency~\cite{Tanford1973,MaibaumDinnerChandler2004}.
When a hydrophobic solute is solvated by water, it excludes water molecules from the region that it occupies, thereby perturbing the inherent interactions between the nearby water molecules 
~\cite{Hummer:1998,ashbaugh_SPT,Chandler:Nature:2005,ball08,berne_rev09,Jamadagni:ARCB:2011,Chandler:2012}.
Because water molecules have strong hydrogen bond interactions, 
it is the penalty for this disruption of the water structure 
 that dominates hydrophobic hydration free energies. 
Hard sphere solutes that simply exclude water molecules from the region they occupy have thus served as 
idealized hydrophobic solutes, and their hydration and assembly have been extensively studied using both molecular simulations~\cite{WidomScience,FHS:1973,CrooksChandler1997,HGC,Huang:JPCB:2002,garde07,garde08,ball08,Patel:JPCB:2010,Patel:JSP:2011,Remsing:2013,Patel:JPCB:2014} and theory~\cite{FHS:1973,Pratt:JCP:1977,Chandler:PRE:1993,ashbaugh_SPT,Hummer:PNAS:1996,Garde:PRL:1996,LCW,weeks_review,berne_rev09,LLCW}.

The use of hard solutes as ideal hydrophobes is particularly judicious, because their excess hydration free energy, $\Delta G_{\rm HS}$,
is intimately tied to fluctuations in water density through~\cite{Widom:JCP:1963,Hummer:PNAS:1996,pratt_book}
\begin{equation}
\beta\Delta G_{\rm HS} = -\ln P_{v}(N=0),
\label{eq:pdt}
\end{equation}
where $P_v(N)$ is the probability of observing $N$ waters in the volume $v$ corresponding to the size and shape of the hard solute, and
$\beta=1/\kT$ ($k_{\rm B}$ is Boltzmann's constant and $T$ is the temperature).
Thus, an understanding of density fluctuations in bulk water has the potential to inform free energies of hydrophobic hydration and association.
Indeed, the notion that small fluctuations in water density obey Gaussian statistics lies at the heart of the Pratt-Chandler theory of hydrophobic hydration~\cite{Pratt:JCP:1977,Chandler:PRE:1993}.
Using molecular simulations, Hummer \emph{et al.} explicitly showed that density fluctuations in small observation volumes ($v\lesssim 1$~nm$^3$) are Gaussian, allowing them to relate $\Delta G_{\rm HS}$ to the moments of $P_v(N)$~\cite{Hummer:PNAS:1996, Garde:PRL:1996}.
This simplification has afforded molecular insights into various phenomena, including entropy convergence, wherein protein unfolding entropies converge at a common temperature~\cite{Garde:PRL:1996,Remsing:2013}, as well as the denaturation of proteins~\cite{tW_rev,pratt_rev,Hummer:PNAS:1998}
and the formation of clathrate hydrates~\cite{Hummer:PNAS:1998} at high pressures.

Noting that water at ambient conditions is in close proximity to liquid-vapor coexistence, the theory of Lum, Chandler, and Weeks (LCW) predicted~\cite{LCW,HuangChandlerPRE,LLCW} that low-$N$ fluctuations should be enhanced (relative to Gaussian statistics) in large volumes ($v\gtrsim 1$~nm$^3$). 
Indeed, simulations subsequently verified that while $P_v(N)$ remains Gaussian near its mean, the likelihood of low-$N$ fluctuations in large $v$ is enhanced significantly in bulk water and even more so near hydrophobic surfaces; that is, $P_v(N)$ develops fat low-$N$ tails~\cite{HuangChandlerPRE,mittal_pnas08,garde09prl,Godawat:PNAS:2009,Patel:JPCB:2010,Acharya:Faraday:2010,Patel:JSP:2011,Patel:PNAS:2011,Patel:JPCB:2012,Jamadagni:ARCB:2011}.
This perspective clarified that water near hydrophobic surfaces sits at the edge of a dewetting transition that can be readily triggered by small perturbations~\cite{Patel:JPCB:2012,berne04,berne05_melittin,chou_dewet,chou05,garde_rev}.
It also led to the prediction that the assembly of small hydrophobic solutes in the vicinity of extended hydrophobic surfaces would be barrierless~\cite{Patel:PNAS:2011,Vembanur:2013}.
Thus, an understanding of density fluctuations in small and large volumes as well as in the vicinity of interfaces has made significant contributions to our understanding of hydrophobic hydration and assembly.

In addition to excluding water, realistic solutes also possess favorable attractive (van der Waals) interactions with water.
Here, our goal is to establish a connection between water density fluctuations and hydration free energies of {\it realistic} solutes
of various sizes in bulk water and at interfaces.
To accomplish this, we first turn on solute-water attractive interactions, and then characterize water density 
fluctuations corresponding to the emptying of the solute repulsive core in the presence of these attractions.
The hydration free energy of realistic van der Waals solutes, $\Delta G_{\rm vdW}$, is then given by:
\begin{equation}
\beta\Delta G_{\rm vdW} = \beta\Delta G_{\rm att} -\ln P_v^{({\rm att})}\para{N=0},
\label{eq:pt2}
\end{equation}
where $\Delta G_{\rm att}$ is the free energy for turning on the solute-water attractions in bulk water,
and $P_v^{({\rm att})}(N)$ is now the probability of observing $N$ waters in $v$ in the presence of the attractive field.
We show that $\Delta G_{\rm att}$ is accurately described by linear response theory, so that an understanding of $P_v^{({\rm att})}(N)$ informs $\Delta G_{\rm vdW}$ in much the same way as $P_v(N)$ has informed $\Delta G_{\rm HS}$.
%

We quantify $P_v^{({\rm att})}(N)$ for spherical volumes of different sizes and a range of attractive strengths.
We find that the presence of attractions in both the solute core, $v$, and its hydration shell, following the Weeks-Chandler-Andersen (WCA) prescription~\cite{wca}, significantly alters water density fluctuations; it becomes progressively harder to empty $v$ as the strength of attractions is increased.
However, as far as the estimation of $\Delta G_{\rm vdW}$ is concerned, this relative difficulty in emptying $v$ in the presence of attractions is largely offset by the favorable free energy, $\Delta G_{\rm att}$, of turning on those attractions in the first place.
To minimize cancellation between the favorable $\beta\Delta G_{\rm att}$ and the unfavorable $-\ln P_v^{({\rm att})}(N=0)$ terms in Equation~\ref{eq:pt2}, we consider an alternative prescription for hydrating the same solute; one that involves turning on attractions in the hydration shell, but not in the solute core, $v$.
The overall value of $\Delta G_{\rm vdW}$ does not depend on which prescription is used and in particular, whether attractions are turned on in the solute core or not; attractions in the core simply increase the magnitude of the components of $\Delta G_{\rm vdW}$.
In the presence of attractions in the hydration shell alone, we find that the water density fluctuations in the core are remarkably unaltered;  attractions effect only a subtle change in the mean density.
We find this to be true for solutes of all sizes and reasonable attraction strengths.
Thus, our primary finding is that water density fluctuations that are relevant to hydrophobic hydration are largely unaffected by the presence of attractions.

Our results also suggest strategies for circumventing the breakdown of perturbation theories of hydrophobic hydration, which occurs for large ($\gtrsim 1$~nm$^3$) solutes~\cite{Underwood:2011,chou05jacs,chou_dewet}.
$\Delta G_{\rm vdW}$ is typically estimated by first creating a cavity and subsequently turning on attractions.
The free energy for turning on attractions can be readily estimated if accurate approximations are available for the response of the hydration shell water density to the attractions.
This approach works well for small solutes because water structure around the cavity is not significantly altered when attractions are turned on, so that water density responds linearly to attractions. 
In contrast, a soft vapor-liquid interface is nucleated around large cavities, and is readily displaced towards the solute when attractions are turned on.
As a result, water density near the cavity increases in a sigmoidal fashion (and not linearly) as the strength of attractions is increased, thereby violating a key assumption that underpins perturbation theory.
Because we turn on attractions in bulk water prior to the formation of a cavity, water density responds linearly to the strength of attractions, and $\Delta G_{\rm att}$ is readily estimated using linear response theory.
Water density fluctuations are largely unaltered when attractions are turned on in the hydration shell alone,
so that $-\ln P_v^{({\rm att})}(N=0)$, and therefore $\Delta G_{\rm vdW}$, can be estimated from a knowledge of water density fluctuations in the absence of attractions.

Furthermore, recognizing that the presence of attractions in the core of a solute has no bearing on its hydration free energy, also allows for efficient estimation of hydration free energies in interfacial environments.
When hydrating a solute in such inhomogeneous environments, attractions from neighboring solutes or interfaces acting on $v$, make it harder to empty the solute core.
We recommend that such attractions should be turned off (at least partially) in a first step, with the corresponding free energy readily estimated using linear response theory. 
The solute core can then be emptied more easily in a second step because the free energy for doing so is substantially reduced.

\section{Methods}

\subsection{Models}

\begin{figure}
\begin{center}
\includegraphics[width=0.5\textwidth]{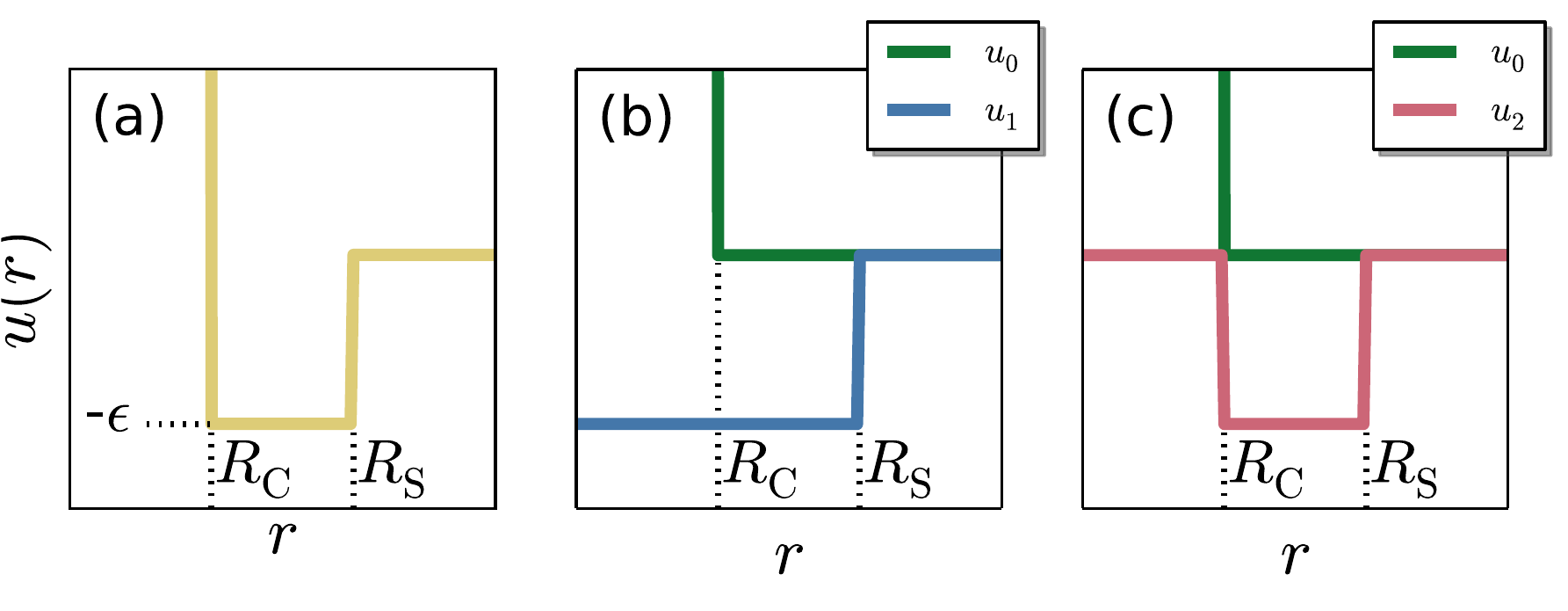}
\end{center}
\vspace{-0.5cm}
\caption[Probe volume sketch]
{(a) The square well potential, $u(r)$, with a hard core repulsion at $r=R_{\rm C}$ and a well of depth, $\epsilon$, and width,
$R_{\rm S}-R_{\rm C}$.
(b) This solute-water potential can readily be split into its repulsive, $u_0(r)$, and attractive, $u_1(r)$, components using a WCA-like prescription~\cite{wca}.
(c) Alternatively, $u(r)$ can be split into $u_0(r)$, and the attractive potential, $u_2(r)$, which is non-zero only in the shell region, $R_{\rm C}<r<R_{\rm S}$.
}
\label{fig:pvsketch}
\end{figure}

To obtain a qualitative understanding of the influence of attractive potentials on density
fluctuations, we focus on the solvation of a square well particle in the extended simple point charge (SPC/E) model of water~\cite{SPCE}.
While we focus on square-well solutes, our findings are general and hold for more realistic Lennard-Jones (LJ) solutes (see Supplementary Material~\cite{SI}).
We chose the SPC/E model for water because it reasonably mimics the bulk density, isothermal compressibility, surface tension, and liquid-vapor coexistence properties of real water; these properties primarily influence the behavior of water density fluctuations~\cite{Chandler:2012}.

{\it Bulk Hydration}:
The solutes we study interact with water oxygens through the pair potential shown in Figure~\ref{fig:pvsketch}a.
The potential has a hard core of radius $R_{\rm C}$ and an attractive region of width $R_{\rm S}-R_{\rm C}$ and depth $\epsilon$.
We consider three core sizes spanning the small to large solute size regimes
of hydrophobic solvation~\cite{LCW}, $R_{\rm C}=0.336$~nm, $R_{\rm C}=0.6$~nm, and $R_{\rm C}=0.9$~nm.
The smallest solute corresponds to the effective hard sphere radius~\cite{Blip,Huang:JPCB:2002} of a united atom methane with
Lennard-Jones parameters, $\sigma_{\rm LJ}=0.3448$~nm and $\epsilon_{\rm LJ}=0.8956$~kJ/mol~\cite{MethaneParms}.
For all solutes, we choose a spherical shell region 0.3~nm in width, that is, $R_{\rm S}=R_{\rm C}+0.3$~nm, and set the well-depth, $\beta\epsilon=1$.
%

While the repulsive part of the square-well potential is unambiguously given by the hard sphere pair potential, $u_0$, the attractive component can be defined in one of two ways (Figure~\ref{fig:pvsketch}).
The first solute-water attractive potential, $u_1(r)$, arises from a WCA-like separation of a standard square-well potential~\cite{wca} shown in Figure~\ref{fig:pvsketch}b, and acts on both the core and shell regions.
The contribution of such solute-water attractions to the Hamiltonian of the system is
\begin{equation}
U_{1}(\Rbar)=-\epsilon(\NC+\NS),
\end{equation}
and depends only on the number of molecules in the core and shell regions, $\NC$ and $\NS$ respectively, where $\Rbar$ denotes a configuration vector containing the positions of all the water oxygens.
The results obtained from this potential can be qualitatively compared to the WCA attractive portion of the LJ potential; such a comparison is shown in the Supplementary Material~\cite{SI}.
Alternatively, an attractive potential that acts on the shell alone, $u_2(r)$, can be considered (Figure~\ref{fig:pvsketch}c), and its contribution to the system Hamiltonian is
\begin{equation}
U_2(\Rbar) = -\epsilon\NS.
\end{equation}
To vary the strength of attractions, we employ a linear coupling parameter in both cases, that is, we apply attractive potentials, $\lambda_i U_i$, and vary $\lambda_i$ to change the interaction strength.
{\it Hydration at Interfaces}:
We also consider the hydration of a hard solute in the vicinity of a square-well attractive surface.
The surface excludes waters from a cuboid-shaped volume, $v_{\mathrm{s}}=2.5\times2.5\times1.0$~nm$^3$, and has attractive interactions of strength $\lambda_1\epsilon$ with water in an adjoining volume, $v_1=2.5\times2.5\times0.3$~nm$^3$.
The cuboid-shaped hard solute, $v=1.5\times1.5\times0.3$~nm$^3$, is then hydrated at the center of $v_1$.

\subsection{Simulation Details and Methods}
All simulation data presented here were obtained using version 4.5.3 of the GROMACS
molecular dynamics (MD) simulation package~\cite{gmx4ref}, modified to include the various external and biasing potentials
used in this work. 
MD simulations were performed in the isothermal-isobaric (NPT) ensemble,
where the canonical velocity-rescaling thermostat of Bussi~\etal~\cite{Bussi:JCP:2007} was used to maintain a constant temperature of 300~K and a Parinello-Rahman
barostat~\cite{Parrinello-Rahman} was used to maintain a pressure of 1~bar. 
Electrostatic interactions were handled
with the particle mesh Ewald method~\cite{PME} with a real space cutoff of 1~nm and grid spacing of 0.12~nm.
Similarly, van der Waals interactions were cutoff at a distance of 1~nm, and standard energy and pressure corrections were used to account for the influence of the truncated interactions~\cite{CompSimLiqs}.

{\it Bulk Hydration}: We denote the free energy of turning on the attractive field, $\lambda_i U_i$ ($i=1,2$), by $\Delta G_i$, and the fluctuations in the presence of the field by $P^{(i)}_v(N)$.
For the specific case of square-well type potentials, the fields, $U_1$ and $U_2$, depend only on $N_v$ and $N_{v_{\rm sh}}$, the number of water molecules in the solute core and hydration shell, respectively;
that is $U_i(\Rbar) = U_i(N_v(\Rbar), N_{v_{\rm sh}}(\Rbar))$. 
Thus, both $\Delta G_i$ and $P^{(i)}_v(N)$ can be readily obtained for a range of attractive strengths, $\lambda_i$, if the probability, $\PNCNS$, of observing $N$ and $N_{\rm sh}$ waters in $v$ and $v_{\rm sh}$, respectively, in the absence of any solute-water attractions is known.
In particular, $P^{(i)}_v(N)$ can be obtained from the exact expression
\begin{equation}
P^{(i)}_v(N) = C\sum _{ {N_{\rm sh}}=0 }^\infty  \PNCNS e^{-\beta \lambda_i U_i(N,N_{\rm sh})},
\label{eq:pivn}
\end{equation}
where $C$ is a normalization constant.
Similarly, $\Delta G_i$ can be determined exactly from
\begin{equation}
\beta\Delta G_i = -\ln \sum_{N} \sum_{N_{\rm sh}} \PNCNS e^{-\beta \lambda_i U_i(N,N_{\rm sh})}.
\label{eq:exactattr}
\end{equation}
Derivations of both Equation~\ref{eq:pivn} and Equation~\ref{eq:exactattr} are included in the Supplementary Material~\cite{SI}.
To obtain $\PNCNS$, we employed indirect umbrella sampling (INDUS)~\cite{Patel:JSP:2011} with harmonic biasing potentials of the form
\begin{equation}
U_{\rm bias}(\Rbar)=\frac{\kappa_1}{2}(\NCt-\tilde{N}^*)^2 +\frac{\kappa_2}{2}(\NSt-\tilde{N}_{\rm sh}^*)^2.
\label{eq:2dindus}
\end{equation}
The INDUS approach dictates coarse-graining the discrete variables $N_v$ and $N_{v_{\rm sh}}$ to the continuous variables $\tilde{N}_v$ and $\tilde{N}_{v_{\rm sh}}$ through convolution
with a truncated and shifted Gaussian; here we use a width of 0.01~nm and a cutoff of 0.03~nm for this smoothing function.
The force constants, $\kappa_i$, and the biased potential minima, $\tilde{N_i}^*$, were tuned to achieve sufficient overlap between neighboring windows.
The corresponding $\PNCNS$ distributions were then obtained from the biased simulations using the unbinned weighted histogram analysis method (UWHAM)~\cite{UWHAM,MBAR}.

{\it Hydration at Interfaces}:  
Because simulating the square-well surface using MD would result in impulsive forces, we instead employ closely related continuous potentials to mimic the square-well potential, and make use of standard reweighting techniques~\cite{Torrie:1977} to estimate the behavior of a true square-well surface.
The exclusion of water molecules from the square-well surface is accomplished by applying the potential
$\tilde{U}_{\rm s}(\Rbar)=\phi_{\rm s}\tilde{N}_{v_{\rm s}}(\Rbar)$,
where $\tilde{N}_{v_{\rm s}}(\Rbar)$ is the coarse-grained number of waters in $v_{\rm s}$, and
$\beta\phi_{\rm s}=12$ was chosen to yield a substantial number of configurations with the volume empty.
Analogously, the attractive potential in the adjoining hydration shell volume, $v_1$, is 
$\tilde{U}_1(\Rbar) = \phi_1 \tilde{N}_{v_1}(\Rbar)$, where $\tilde{N}_{v_1}$ is the coarse-grained number of water molecules in $v_1$, and $\phi_1 \equiv -\lambda_1\epsilon$.
Both $\tilde{N}_{v_s}$ and $\tilde{N}_{v_1}$, were defined according to the INDUS prescription described above.
To estimate the probability, $P^{(1)}_v(N)$, of observing $N$ water molecules in $v$, a series of INDUS simulations were then performed using a biasing potential similar to the one in Equation~\ref{eq:2dindus}, but with $\kappa_2=0$.
Averages in a system with a true square-well surface, $\avg{(\cdots)}$, can be related to averages in the mimic square-well system, $\avg{(\cdots)}'$ (simulated using the potentials $U_s$ and $U_1$), through reweighting as,
\begin{equation}
\avg{(\cdots)} = \frac{ \avg{ (\cdots) \delta_{0,N_{v_{\rm s}}(\Rbar)} e^{ \beta\phi_{\rm s}\tilde{N}_{v_{\rm s}}(\Rbar) - \beta\phi_1 [ N_{v_1}(\Rbar)-\tilde{N}_{v_1}(\Rbar) ] }}'} { \avg{\delta_{0,N_{v_{\rm s}}(\Rbar)}e^{ \beta\phi_{\rm s}\tilde{N}_{v_{\rm s}}(\Rbar) - \beta\phi_1 [ N_{v_1}(\Rbar)-\tilde{N}_{v_1}(\Rbar) ] }}'},
\end{equation}
where $\delta_{i,j}$ is the Kronecker delta function, and $N_{v_{\rm s}}$ and $N_{v_1}$ are the number of waters in $v_{\rm s}$ and $v_1$ respectively.
This average, in turn, is evaluated using UWHAM~\cite{UWHAM,MBAR}.
The probability, $P^{(2)}_v(N)$, of finding $N$ waters in $v$, in the absence of favorable surface-water interactions in $v$, 
is obtained through another reweighting,
\begin{equation}
P^{(2)}_v(N) = C' P^{(1)}_v(N) e^{-\beta\lambda_1\epsilon N},
\end{equation}
where $C'$ is a normalization constant.

\section{The Influence of Attractions on Water Density Fluctuations}

\begin{figure}
\begin{center}
\includegraphics[width=0.5\textwidth]{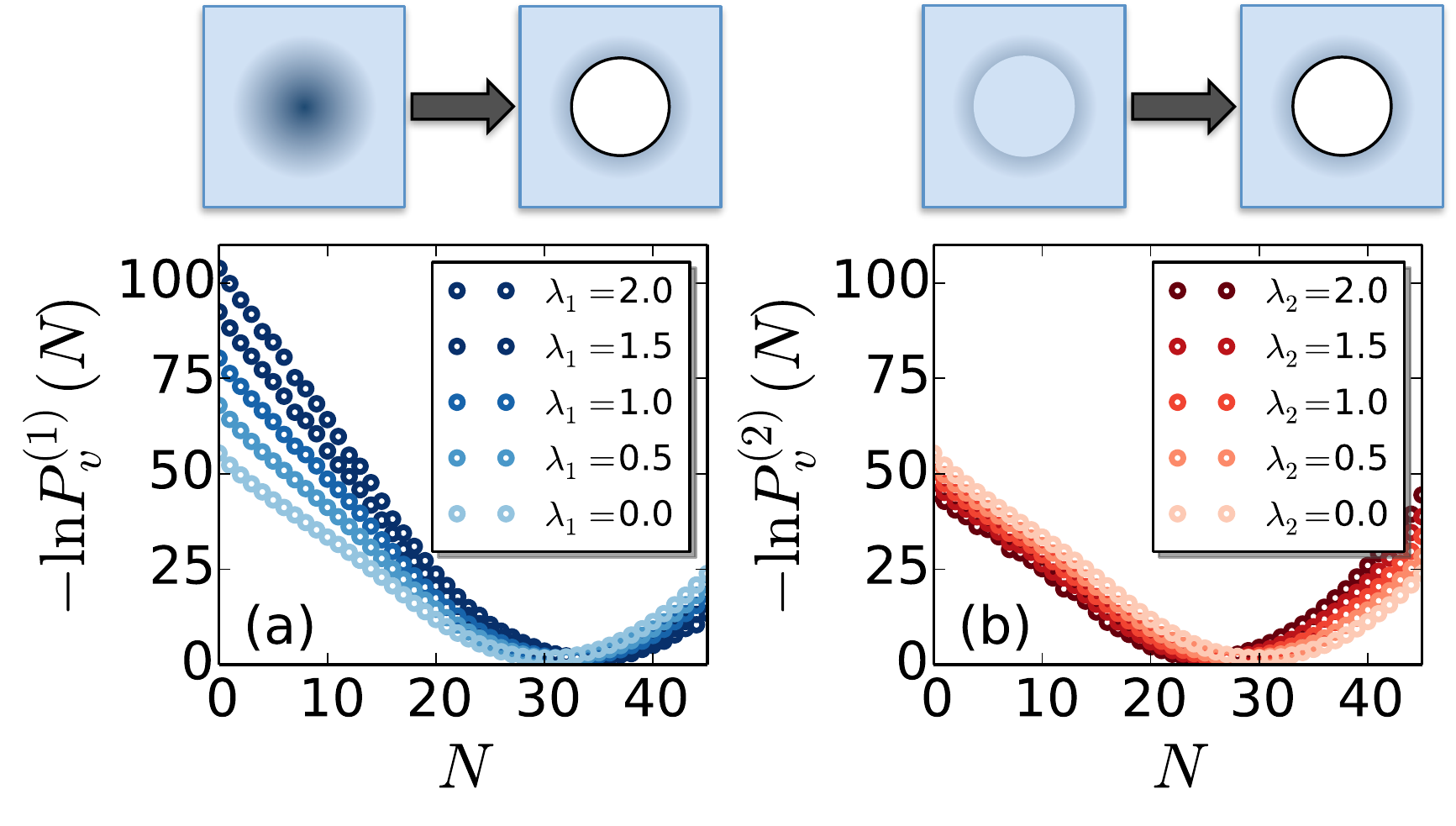}
\end{center}
\vspace{-0.5cm}
\caption[Density Fluctuations]
{(a) The presence of an attractive field that acts on \emph{both} the core and shell regions, $\lambda_1 U_1$, 
drastically alters density fluctuations in the core region, as shown here for a spherical solute with $R_{\rm C}=0.6$~nm and $R_{\rm S}=0.9$~nm.
On increasing the attractive strength $\lambda_1$, the tails at low $N$ become steeper, indicating that it is more difficult to evacuate the core volume.
(b) In contrast, fluctuations in the core region are not significantly perturbed by the attractive field, $\lambda_2 U_2$, which acts on the hydration shell alone.
As the strength of the attractive interactions, $\lambda_{2}$, is increased, only a small \emph{decrease} in the average number of molecules in the core is observed.}
\label{fig:densflucs}
\end{figure}

%
The attractive field, $\lambda_1 U_1$, which acts on both the core and shell regions, significantly affects water density fluctuations in the core, as shown in Figure~\ref{fig:densflucs}a for $R_{\rm C}=0.6$~nm.
As expected, the average number of molecules in the core region, $\avg{\NC}_{\lambda_1}$, increases with increasing
field strength $\lambda_1$.
Here, $\avg{(\cdots)}_{\lambda_i}$ represents an ensemble average in the presence of the attractive field $\lambda_i U_i$.
Importantly, the non-Gaussian nature, or the fatness of the low-$N$ tails is also diminished, making it increasingly difficult to empty the core region as $\lambda_1$ is increased.
%
\begin{figure*}
\begin{center}
\includegraphics[width=0.8\textwidth]{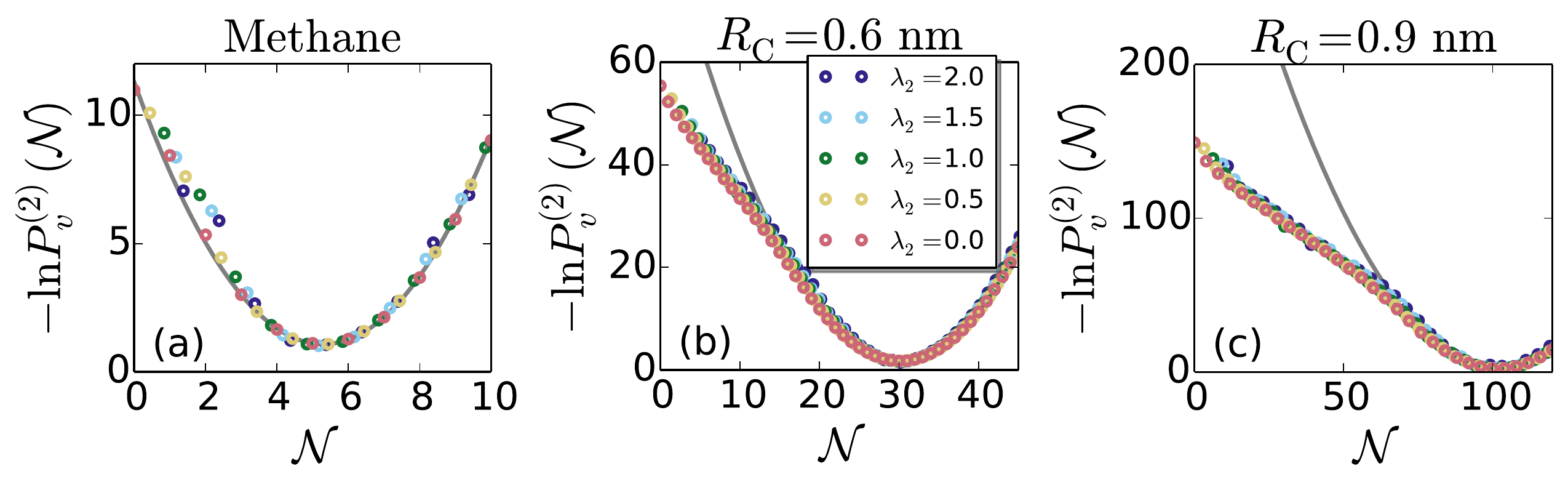}
\end{center}
\vspace{-0.5cm}
\caption[Shell Attractions Decrease Average Number of Waters in Core]
{
Water number distributions, $P_v^{(2)}$, in the presence of attractive fields $\lambda_2 U_2$, shifted horizontally to align their means, $\avg{\NC}_{\lambda_{2}}$, are shown for probe volumes with radii (a) equivalent to that of a methane, (b) 0.6~nm, and (c) 0.9~nm.
Fluctuation spectra obtained over a range of attractions collapse onto a single curve when plotted as a function of $\mathcal{N}\equiv N - \Delta N$, where $\Delta N\equiv\avg{\NC}_{\lambda_2}-\avg{\NC}_0$, highlighting that the presence of an attractive potential in the hydration shell simply decreases the average number of water molecules in the core, and not the fluctuations around the mean.
This is true independent of solute size and whether the fluctuations follow Gaussian statistics (solid gray lines).
}
\label{fig:figure3}
\end{figure*}
%
%
To better understand how water density fluctuations are affected by $\lambda_1 U_1$, we extend the model for density fluctuations proposed by Huang and Chandler~\cite{HuangChandlerPRE} (See Appendix~\ref{sec:HC}).
The Huang-Chandler model assumes that reducing $N_v$ below its average value results in the formation of a bubble, and that the low-$N$ tail in the water density fluctuations can be described by the energetics of growing the bubble. 
As shown in the Appendix~\ref{sec:HC}, in the presence of $\lambda_1 U_1$, additional work has to be done against the attractive field to expand the bubble, which results in the low-$N$ tails of the $P_v^{(1)}(N)$ distributions being diminished.
This qualitative behavior is also observed for both smaller and larger solutes (See Supplementary Material~\cite{SI}).

In contrast, in the presence of an attractive field such as $\lambda_2 U_2$, which acts only in the hydration shell, the nature of the density fluctuations in the core region is not changed; see Figure~\ref{fig:densflucs}b.
Only a small decrease in the mean, $\avg{\NC}_{\lambda_2}$, is observed when the attractive strength, $\lambda_2$, is increased.
In Figure~\ref{fig:figure3}, the fluctuation spectra, $P_v^{(2)}$, are shifted horizontally so that their means are aligned.
For all the $R_\mathrm{C}$-values studied, the shifted spectra, $P_v^{(2)}(\mathcal{N})=P_v^{(2)}(N-\Delta N)$, where $\Delta N\equiv \avg{\NC}_{\lambda_{2}} - \avg{\NC}_0$, corresponding to various attractions, $\lambda_2$, collapse onto a universal curve.
Remarkably, this invariance of water density fluctuations around the mean is independent of whether the fluctuations are nearly Gaussian (small $R_\mathrm{C}$) or have pronounced fat tails (large $R_\mathrm{C}$), 
This is the central result of this work. 

To understand the basis of this invariance, we recognize that the addition of a linear potential (such as $\lambda_2 N_{v_{\rm sh}}$) to a perfectly harmonic basin (such as one arising from $P_{v,v_{\rm sh}}$ being Gaussian) simply translates the basin (and hence the entire distribution) horizontally~\cite{Patel:JPCB:2012}.
Attractions in the shell favor configurations with more waters than the average, coupling to the high-$N_{\rm sh}$ region of $\PNCNS$. 
Such high-$N$ regions of water number distributions in bulk water have previously been shown to follow Gaussian statistics~\cite{Patel:JPCB:2010,Patel:JSP:2011}, and $\PNCNS$ similarly follows a bivariate normal distribution with nonzero correlation between $N$ and $N_{\rm sh}$ near its mean.
Thus, the $\lambda_2 U_2$ potential, which is linear in $\NS$, and couples to the Gaussian high-$N_{\rm sh}$ region of $\PNCNS$ is expected to shift the entire $\PNCNS$ distribution towards higher $N_{\rm sh}$-values.
As $\lambda_2$ is increased, the correlations between $\NS$ and $\NC$ then alter $\avg{\NC}$, but leave the remainder of $P_v^{(2)}(N)$ unaltered.

\begin{figure}
\begin{center}
\includegraphics[width=0.46\textwidth]{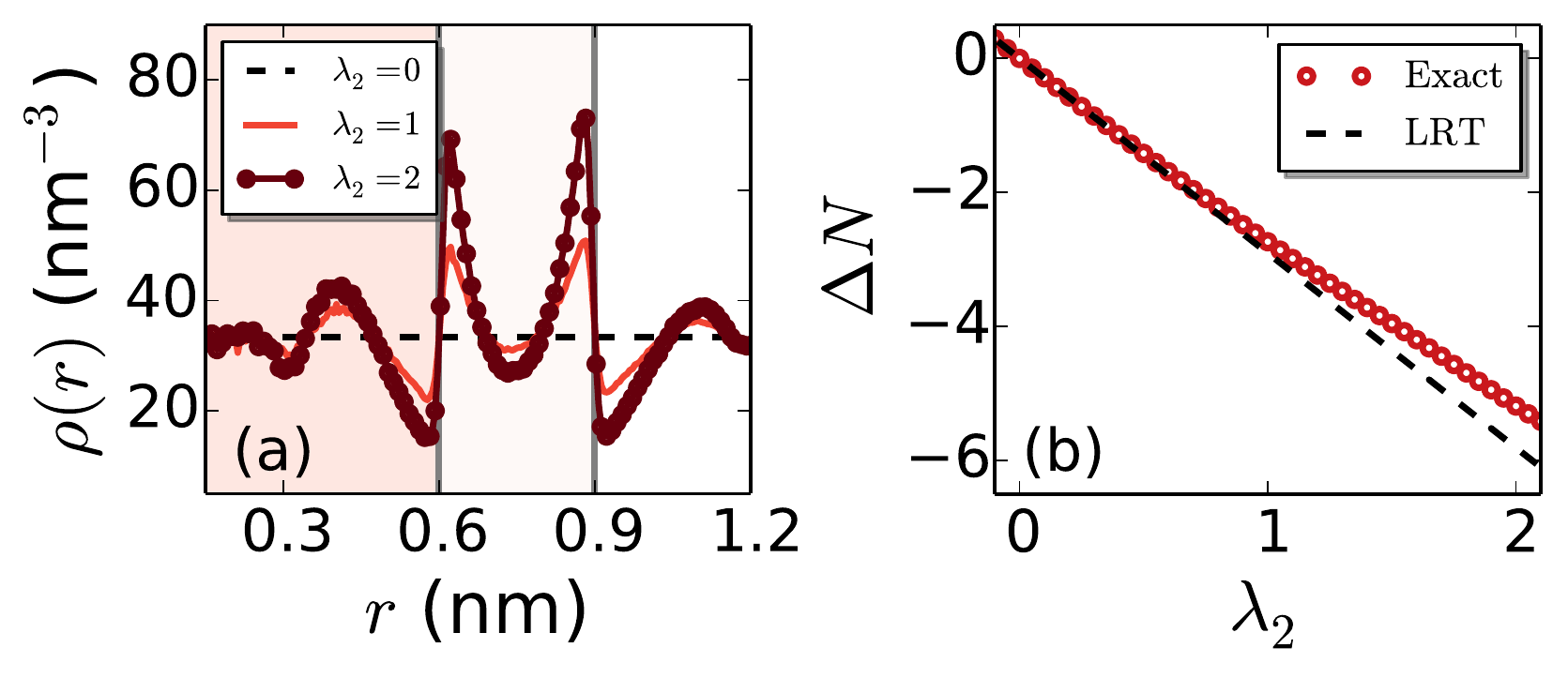}
\end{center}
\vspace{-0.5cm}
\caption[Density Response to Fields]
{
(a) Attractions in the shell region, indicated by vertical gray lines, increase water density in the shell, but result in a decrease in the average water density in the core region, as shown here for  $R_{\rm C}=0.6$~nm. 
The decrease in the core density on increasing $\lambda_2$ is a direct result of layering at the interface between the core and shell regions.
(b) This decerase in the average number of water molecules in the core region, $\Delta N$, as the strength of the potential $\lambda_2$ is increased, can be accurately predicted with linear response theory (LRT), Equation~\ref{eq:LRTN}.
}
\label{fig:density}
\end{figure}

While the application of $U_2$ increases water density in the shell region, packing effects can lead to a concomitant \emph{decrease} in the average number of waters in the core region.
The layering of water density at the core-shell interface, shown in Figure~\ref{fig:density}a, highlights that attractions in the shell indeed decrease the average density in the core.
Interestingly, as shown in Figure~\ref{fig:density}b, this decrease in the average number of molecules in the core region upon application of the potential $U_2$ can be accurately estimated using linear response theory,
\begin{equation}
\Delta N\approx -\lambda_{2}\beta\epsilon\avg{\delta \NC \delta \NS}_0,
\label{eq:LRTN}
\end{equation} 
given the correlation, $\avg{\delta \NC \delta \NS}_0$, between fluctuations in $\NC$ and $\NS$ in the absence of attractions.
As discussed in section V, this agreement will facilitate the development of approximations for estimating hydration free energies of realistic solutes in bulk water.
For now, having quantified the ease with which the core can be emptied in the presence of the attractive fields $\lambda_1 U_1$ and $\lambda_2 U_2$, we turn to the free energies of turning on these potentials in bulk water.
%

\section{Attractions in the Solute Core Do Not Influence Its Hydration Free Energy}

\begin{figure}
\begin{center}
\includegraphics[width=0.48\textwidth]{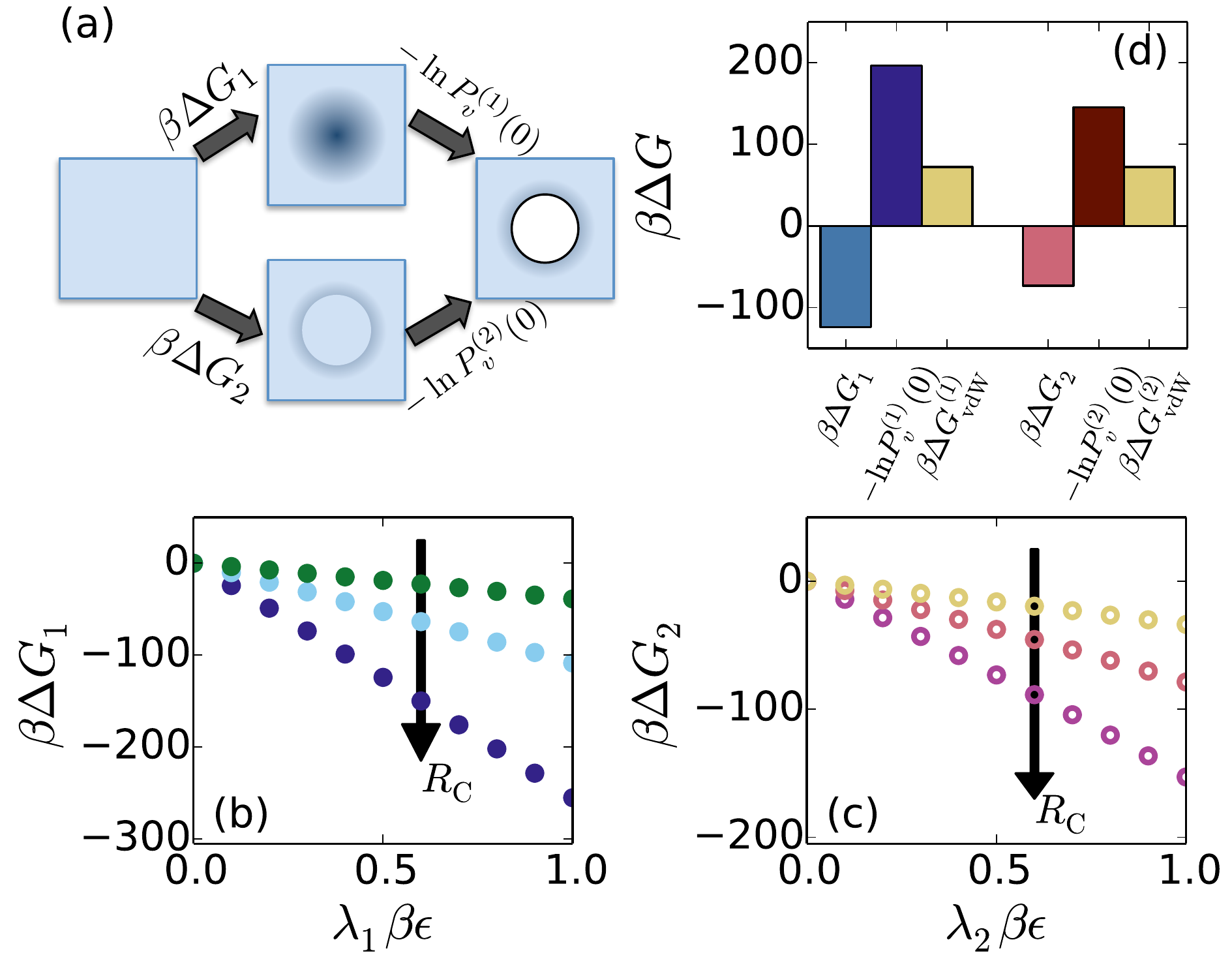}
\end{center}
\vspace{-0.5cm}
\caption[U1 vs U2]
{
(a) The hydration of a realistic hydrophobic solute is carried out in two steps: an attractive field, $\lambda_i U_i$, corresponding to the solute-water attractions is first turned on in bulk water, and water is emptied from the solute core in a subsequent step.
The corresponding free energies are denoted by $\beta\Delta G_i$ and $-\ln P_v^{(i)}(0)$ respectively, where $i=1$ represents attractions in both the core and shell regions (top) and $i=2$ represents attractions in only the shell region (bottom).
(b, c) The free energy of turning on the attractive potential $U_1$ is much larger than that for turning on $U_2$.
(d) This results in a larger cancellation between the favorable $\beta\Delta G_i$ and the unfavorable $-\ln P_v^{(i)}(0)$ free energies for $i=1$, because the total hydration free energy $\Delta G_{\rm vdW} = \Delta G_i -\ln P_v^{(i)}(0)$
is the same for both paths, as shown here for $R_{\rm C}=0.9$~nm and $\lambda_i\beta\epsilon=0.5$.
}
\vspace{-0.5cm}
\label{fig:u1vu2}
\end{figure}

Two alternative paths for hydrating the square-well solute of Figure~\ref{fig:pvsketch}a are shown in Figure~\ref{fig:u1vu2}a.
In the upper path, the attractive potential $\lambda_1 U_1$ is first turned on in the core and the shell, whereas in the lower path, the potential, $\lambda_2 U_2$, is turned on in the shell alone.
We denote the free energy for turning on the attractive field, $\lambda_i U_i$, in bulk water by $\Delta G_i$ ($i=1,2$).
Then, in the second step, a cavity is created in the core region in the presence of the corresponding attractive potential.
The free energy for creating such a cavity is given by $-k_{\mathrm{B}}T\ln P^{(i)}_v(N=0)$, and is informed by the statistics of density fluctuations, $P^{(i)}_v(N)$, shown in Figure~\ref{fig:densflucs}. 

As shown in Figures~\ref{fig:u1vu2}b and c, turning on attractions in the both the core and shell regions results in a large and negative free energetic gain $\Delta G_1$, whereas the corresponding free energy, $\Delta G_2$, for turning on attractions in the shell alone, is significantly smaller in magnitude.
However, the total hydration free energy is given by the sum of the favorable $\Delta G_i$ and unfavorable $-k_{\mathrm{B}}T\ln P^{(i)}_v(N=0)$ terms.
The diminished fluctuations in the presence of $U_1$ (Figure~\ref{fig:densflucs}a) mean that the unfavorable free energy to form a cavity is also larger in the presence of $U_1$.
The total hydration free energy, $\Delta G_{\rm vdW}$, estimated through the upper path thus results in a significant cancellation between the 
 favorable $\Delta G_1$ and the unfavorable $-k_{\mathrm{B}}T\ln P^{(1)}_v(N=0)$ terms, as shown in Figure~\ref{fig:u1vu2}d.
Such large cancellations can lead to significant numerical uncertainty in the estimation of $\Delta G_{\rm vdW}$.

In contrast, the lower path in Figure~\ref{fig:u1vu2}a reduces the extent of such a cancellation by
recognizing that the presence or absence of attractions in the solute core plays no role in determining the overall solute hydration free energy, and minimizing the volume over which the attractions act.
In principle, the cancellation between the favorable $\beta\Delta G_i$ and unfavorable $-\ln P^{(i)}_v(N=0)$ terms can be further reduced or even eliminated by turning on repulsions in the core in conjunction with attractions in the shell in the first step.
Even more complex, low-variance pathways to hydration may also be chosen to cleverly minimize the computational overhead~\cite{Pham:2012aa,Naden:2014}.
However, the advantage of the paths considered here is that they allow for particularly simple analytical approximations for $\Delta G_i$, as discussed in the next section.

\section{Efficient Estimation of Bulk Hydration Free Energies}

\begin{figure}
\begin{center}
\includegraphics[width=0.48\textwidth]{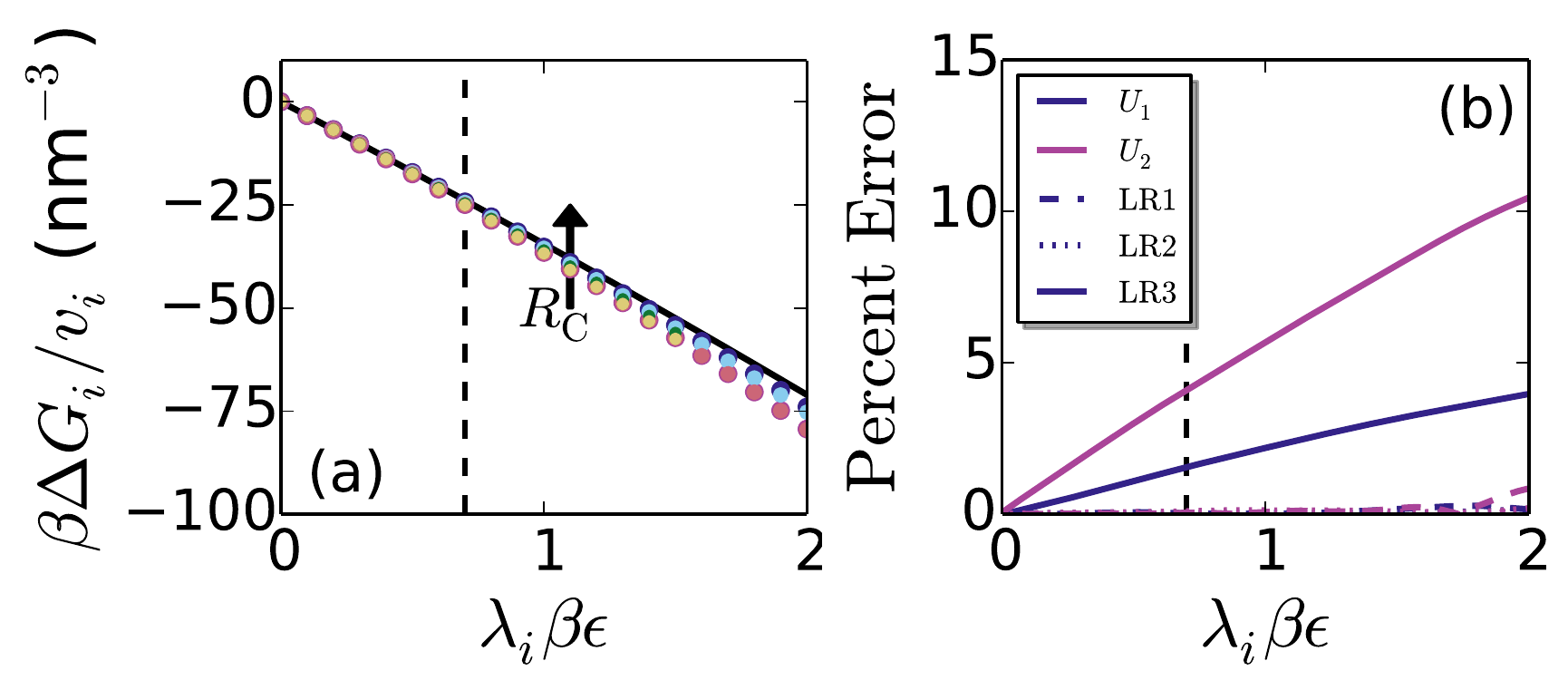}
\end{center}
\vspace{-0.5cm}
\caption[LRT]
{
(a) The free energy, $\Delta G_i$, of turning on the attractive potential, normalized by the volume, $v_i$, on which the attractions act,
is largely independent of solute size and the the range over which the potential acts.
$\Delta G_i$ is well described by the linear response prediction LR3 of Equation~\ref{eq:sqapprox} (solid black line).
(b) The three linear response theory estimates for $\Delta G_i$ are found to be highly accurate,
as shown here for the $R_{\rm C}=0.9$~nm solute.
The dashed line indicates a typical magnitude of attractions for realistic solutes~\cite{fnote_typ-attr}.
}
\vspace{-0.5cm}
\label{fig:u2lrt}
\end{figure}

%
Because attractive fields couple to the Gaussian high-$N$ tails of water density distributions, 
we expect linear response theory to provide an accurate estimate of the free energy for turning on such fields.
In particular, we consider the following approximate forms of Equation~\ref{eq:exactattr}:
\begin{align}
\beta\Delta G_i &\approx \frac{\beta \lambda_i}{2}\brac{\avg{U_i(\Rbar)}_0+\avg{U_i(\Rbar)}_{\lambda_i}} \label{eq:LR1} \\
&\approx \beta\lambda_i\avg{ U_i(\Rbar)}_0 - \frac{\beta^2\lambda_i^2}{2}\avg{(\delta U_i(\Rbar))^2}_0 \label{eq:LR2},
\end{align}
referred to as LR1 and LR2, respectively.
In the context of thermodynamic integration, LR1 represents the application of a trapezoid rule to integrate over the entire range of $\lambda_i$,
while LR2 uses the initial slope of the thermodynamic force at $\lambda_i=0$ and extrapolates over the range of $\lambda_i$ to integrate.
For the square-well potentials considered here, the free energy of turning on
the attractive external field can further be approximated analytically, yielding
\begin{equation}
\frac{\beta\Delta G_i}{v_i} \approx -\rhob(\lambda_i\beta\epsilon)  - \frac{\rhob^2\kappa_T \kT}{2} (\lambda_i\beta\epsilon)^2,
\label{eq:sqapprox}
\end{equation}
where $v_i$ is the volume over which the potential is applied, and $\rhob$ and $\kappa_T$ are the density and isothermal compressibility of bulk water, respectively.
To derive Equation~\ref{eq:sqapprox} (referred to as LR3) from LR2, we recognize that fluctuations in $U_i$ are related to fluctuations in water density, which in turn are related to the isothermal compressibility.
The exact derivations of LR3 and its generic form applicable to slowly-varying potentials are provided in the Supplementary Material~\cite{SI}.

\begin{figure}
\begin{center}
\includegraphics[width=0.48\textwidth]{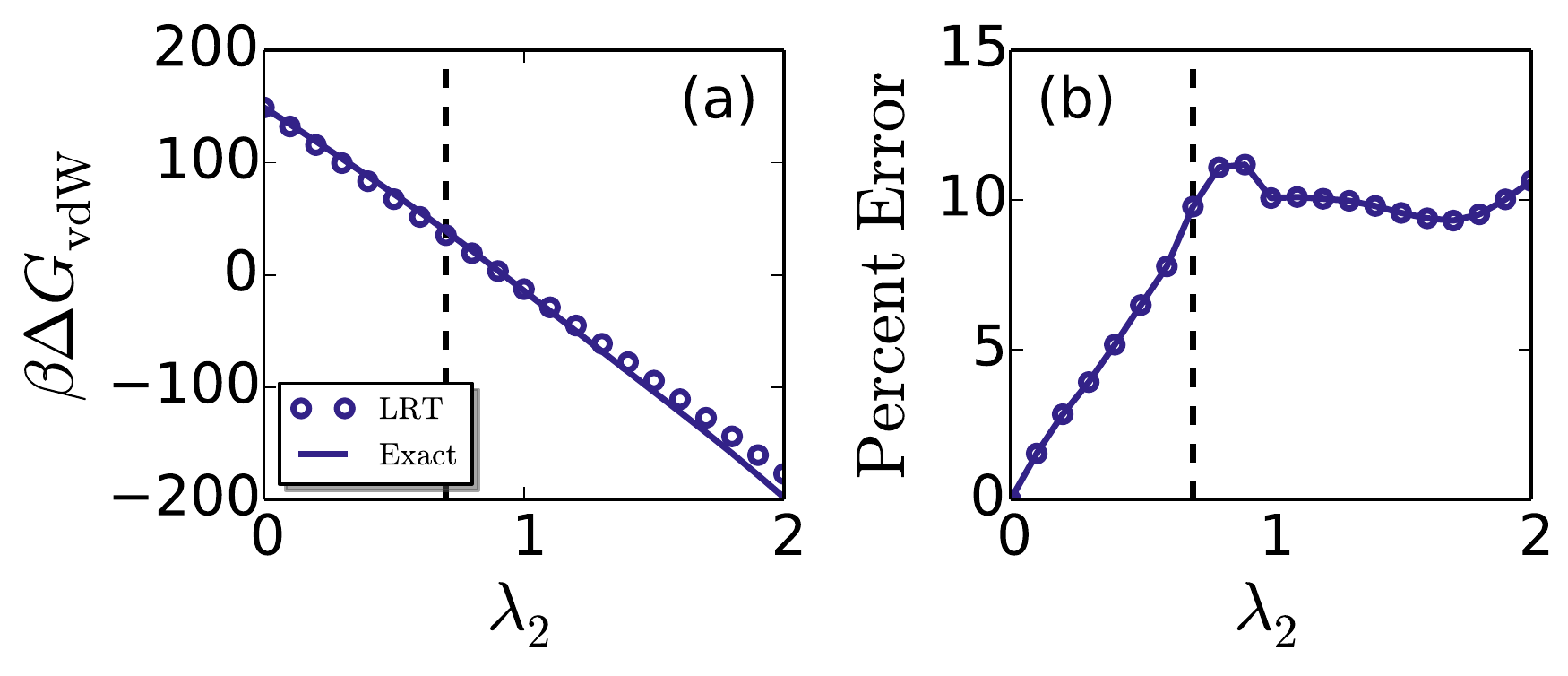}
\end{center}
\vspace{-0.5cm}
\caption[Total FE]
{
(a) The total hydration free energy of a realistic hydrophobic solute, $\Delta G_{\rm vdW}=\Delta G_2 -\kT\ln P_v^{(2)}(N=0)$, can be described with good accuracy
using the approximations in Equations~\ref{eq:sqapprox} and~\ref{eq:pnapprox}, 
as shown here for the solute with $R_{\rm C}=0.9$~nm.
(b) Indeed, for realistic values~\cite{fnote_typ-attr} of $\lambda_2$ (dashed line), the error in $\Delta G_{\rm vdW}$ is less than ten percent.
}
\vspace{-0.5cm}
\label{fig:totalfe}
\end{figure}

LR3 suggests that $\Delta G_i$ can be estimated simply from a knowledge of the strength of the attractions, the volume over which they act, and bulk properties of the solvent.
It predicts that $\Delta G_i$ is proportional to the volume, $v_i$, over which the attractive interactions act, and is quadratic in the strength of the attractions, $\lambda_i$, consistent with previous findings~\cite{Maibaum:2007}. 
We find LR3 to be true to a very good approximation, as illustrated in Figure~\ref{fig:u2lrt}; it begins to break down only for small volumes and large $\lambda_i$ (see Supplementary Material~\cite{SI}).
The discrepancy between the predicted and simulation results for typical attraction strengths~\cite{fnote_typ-attr} is less than five percent.
The approximations, LR1 and LR2, provide even more accurate estimates of $\Delta G_i$, with errors of less than a percent.
In addition to being able to obtain accurate estimates for $\Delta G_i$ from linear response theory, our essential finding that attractions (in the hydration shell of a solute) do not affect water density fluctuations, also permits accurate estimates of the free energy for emptying the solute core, $-\ln P^{(2)}_v(N=0)$.
Turning on attractive interactions in the hydration shell only results in a small shift in the mean, $\Delta N$ (Figure~\ref{fig:figure3}), which itself can be readily estimated using linear response theory (Equation~\ref{eq:LRTN}), so that fluctuations in the presence and absence of attractions can be related through
\begin{equation}
P_v^{(2)}(N)\approx P_v(N+\Delta N).
\label{eq:pnapprox}
\end{equation}
Evaluating the components of the total bulk hydration free energy (Equation~\ref{eq:pt2}) using Equation~\ref{eq:sqapprox} and Equation~\ref{eq:pnapprox} (for $N=0$) leads to accurate predictions for $\Delta G_{\rm vdW}$, as shown in Figure~\ref{fig:totalfe} for the $R_{\rm C}=0.9$~nm solute.
This approach thus presents an efficient way to accurately estimate bulk hydration free energies of realistic attractive solutes, $\Delta G_{\rm vdW}$, starting from information on the statistics of water density fluctuations in bulk water, $P_v(N)$.

\section{Efficient Estimation of Hydration Free Energies in Heterogenous Environments}

\begin{figure*}
\begin{center}
\includegraphics[width=0.9\textwidth]{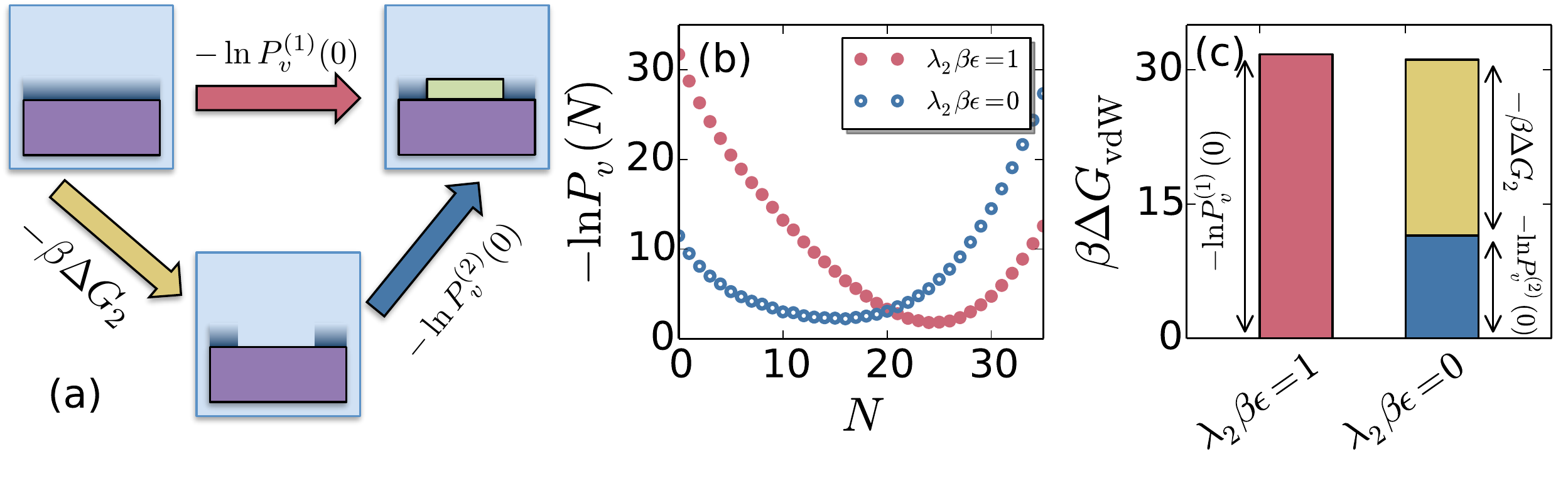}
\end{center}
\vspace{-0.75cm}
\caption[Hard Cube Binding]
{
(a) Top: The hydration of a $1.5\times1.5\times0.3$~nm${^3}$ cuboid-shaped hard solute (green) adjacent to a $2.5\times2.5\times1.0$~nm$^{3}$ attractive square-well surface (purple, well depth $\beta\epsilon=1$ and width $0.3$~nm), entails the creation of a cavity in the presence of attractive surface-water interactions. 
Bottom: Alternatively, the solute can be hydrated in two steps; turning off the attractions in the core of the solute in the first step, followed by creating a cavity in the second step.
(b) Emptying the solute core requires a significantly smaller free energy in the absence of attractions in the core.
(c) While the solute hydration free energy is the same in both cases, the free energy for turning off the attractions, $-\beta\Delta G_{2}$, can be accurately estimated by using linear response theory, and the free energy for subsequently emptying the solute core, $-\ln P_v^{(2)}(0)$, is smaller than $-\ln P_v^{(1)}(0)$, and therefore easier to estimate.
}
\vspace{-0.5cm}
\label{fig:hardcube}
\end{figure*}
%
Figure~\ref{fig:u1vu2}d highlights that it is efficient to turn off any attractive fields in the solute core before estimating the free energy for emptying the core. 
Such attractive fields (acting on water) can originate not only from the solute itself, but also from neighboring solutes or interfaces.
To demonstrate this, in Figure~\ref{fig:hardcube}a, we illustrate the hydration of a hard solute (shown in green) in the vicinity of a surface (shown in purple) that interacts with water through an attractive square well potential (dark blue gradient).
The attractive interactions from the square well act on the core region of the solute, making it harder to empty (upper path).
Alternatively, the attractive potential can first be turned off in the solute core region, as shown in the lower path of Figure~\ref{fig:hardcube}a,
with the corresponding free energy accurately estimated by linear response theory.
Turning off the attractions also ensures that emptying the solute cavity is easier, making it the more efficient alternative (Figure~\ref{fig:hardcube}c).

Indeed, the free energy needed to create a cavity in the presence of the attractive field
is significantly higher (roughly $30~\kT$ versus $10~\kT$), as seen in Figure~\ref{fig:hardcube}b.
When the attractive field is turned off in the solute core,
the average number of water molecules in the core decreases and a significant non-Gaussian tail
emerges in $P_v^{(2)}(N)$  at low $N$; both effects act in concert to lower the free energy needed to form a cavity.
Such a lowering of the cavity formation free energy allows for more efficient estimation of hydration free energies
in heterogeneous environments, because fewer biased simulations are needed to probe the entire
range of density fluctuations in the solute core. 
{\it Caveat}:
While the free energy of turning off the attractive interactions inside the solute core was accurately described using
linear response theory (LR1) here, this may not always be the case.
In particular, turning off attractions completely in a sufficiently large volume adjacent to an extended hydrophobic surface can dewet the volume, causing linear response to break down~\cite{Underwood:2011}.
However, only weak attractions are needed to wet a hydrophobic surface; recent work suggests that an attractive strength of roughly $0.1$~\kT~\cite{Godawat:2014,Willard:2014} is sufficient to prevent dewetting. 
Thus, attractions may be turned off partially without triggering dewetting, and the corresponding free energy safely estimated using linear response.
In such cases, our revised recommendation is to reduce the surface-water attractions down to $0.1$~\kT \ in a first step, followed by cavity creation in the presence of reduced attractions.

\section{Conclusions}

Water number fluctuations in small volumes follow Gaussian statistics, while
those in large volumes have non-Gaussian tails at low density.
An understanding of these fluctuations has made seminal contributions to our understanding of the hydration and interactions of idealized, purely repulsive hydrophobic solutes.
To similarly inform the hydration of realistic hydrophobic solutes that have attractive interactions with water, here we have provided a description of these fluctuations in the presence of attractive fields.
We find that WCA-like attractive potentials~\cite{wca}, which are non-zero both inside and outside the solute core, alter the nature of water density fluctuations significantly, making it more difficult to create a cavity.
In contrast, when attractions are turned on in the hydration shell alone, water density fluctuations are largely unaltered in both small and large volumes.

We also find that the favorable free energy for turning on the various attractive fields is readily approximated by linear response theory and is proportional to the volume on which the fields act.
Consequently, the free energy of turning on attractions in both the core and the shell is much larger than the free energy of turning on attractions in the hydration shell alone.
When attractions are turned on in both the core and the shell, the favorable free energy of turning on those attractions and the unfavorable free energetics of emptying the solute core in the presence of those attractions, largely cancel in the estimation of the total solute hydration free energy.
This cancelation is minimized when the solute-solvent attractions are turned on in the hydration shell alone, making it the more efficient route for estimating hydration free energies of realistic solutes.

In addition to informing hydrophobic hydration and interactions, an understanding of density fluctuations 
has also facilitated the development of novel simulation methods. 
By leveraging an understanding of water density fluctuations, and the associated response of water density to perturbations, Patel and Garde recently introduced a method for estimating cavity hydration free energies that is two orders of magnitude more efficient as compared with umbrella sampling~\cite{Patel:JPCB:2014}.
Our characterization of density fluctuations in the presence of slowly-varying attractive interactions has similarly allowed us to suggest strategies for more efficiently estimating hydration free energies of realistic solutes, both in bulk water and in the vicinity of interfaces.

Our approach involves turning on attractions in bulk water, followed by emptying the solute core, and is in contrast with traditional methods that first create a cavity in bulk water, and subsequently turn on attractions. 
Because attractions have little effect on the structure of water adjacent to small cavities (small enough for the hydrogen bond network of water to be maintained around them)~\cite{Huang:JPCB:2002,Maibaum:2007,Galamba:2013,Remsing:2013}, traditional methods can 
readily estimate the free energy for turning on attractions using linear response theory.
However, larger cavities induce dewetting in their vicinity; attractions rewet the solute, so that water density and consequently the solute-water interaction energy do not vary linearly with the strength of attractions~\cite{Chandler:Nature:2005}.
Indeed, recent work from Underwood and Ben-Amotz~\cite{Underwood:2011} has shown a transition from linear to non-linear response as the solute size is increased~\cite{Underwood:2011}, occurring at roughly 1~nm; the length scale corresponding to the crossover in hydrophobic hydration~\cite{LCW}.
Because our approach involves turning on attractions in bulk water \emph{before} creating a solute core, it circumvents the breakdown of linear response theory that plagues the traditional perturbative treatment of attractive interactions. 

The approach to solvation presented here is similar in spirit to the two-step method considered by Weeks and coworkers~\cite{Weeks:1998,weeks_review}
in their development of the molecular-scale van der Waals (MVDW) theory for inhomogenous systems,
which was subsequently combined with the Gaussian field model of Chandler~\cite{Chandler:PRE:1993} to yield the LCW theory of hydrophobicity~\cite{LCW}.
The MVDW theory first considers turning on all slowly-varying interactions in the bulk fluid, which include not only the solute-solvent attractions that we consider here, but also long-ranged \textit{solvent-solvent} interactions.
Near large hydrophobic solutes, the long-ranged solvent-solvent attractive interactions are unbalanced and result in a net repulsion away from the solute~\cite{weeks_review}.
Such repulsions couple to the non-Gaussian low $N$ tails in water number distributions and result in dewetting; thus their effect can not be captured using linear response.
Recognizing this, the LCW~\cite{LCW} theory employed a Landau-Ginzburg free energy functional that enables repulsive potentials to induce dewetting.
In contrast, because our focus here has been on slowly-varying attractive potentials that couple to the Gaussian high-$N$ region of water number distributions, we are able to make judicious use of linear response theory.
The results presented here rely on two essential properties of attractive interactions: (i) they are slowly varying and thereby minimally perturb the structure of water, and (ii) they couple to Gaussian tails of water number distributions.
Thus, the lessons learned from this work will also be applicable to other interactions that possess these two characteristic features.
In particular, recent work by Weeks and co-workers~\cite{Chen:2004,Chen:2006,Rodgers:2008,LMFWater,Remsing:2011,Remsing:2013} has shown that it is instructive to resolve electrostatic interactions into short- and long-ranged components, in analogy with the WCA prescription, which divides the Lennard-Jones potential into a short-ranged repulsive and a slowly-varying attractive component~\cite{wca}.
While the long-ranged electrostatic component follows linear response theory, underpinning dielectric continuum theories~\cite{RemsingThesis}, 
the short-ranged interactions are more complex, leading to directional hydrogen bonds and specific ion effects~\cite{BeckLengthscales,Collins:2012}.
Thus, in hydrating ions, for example, it would be instructive to turn on the long-ranged component of electrostatic interactions first, and investigate how it affects water density fluctuations; such calculations will be the focus of future work.

\begin{acknowledgements}
The authors acknowledge partial financial support from the National Science Foundation through a Seed Grant from the University of Pennsylvania Materials Research Science and Engineering Center (NSF UPENN MRSEC DMR 11-20901).
\end{acknowledgements}

\appendix


\section{Extension of the Huang-Chandler Model}
\label{sec:HC}

Here, we describe the response of $P_v^{(1)}(N)$ to an attractive potential by extending the model developed by Huang and Chandler (HC)~\cite{HuangChandlerPRE}.
The model assumes that reducing $\NC$ below its average value results in the formation of a spherical bubble of radius $r_b$,
and approximates the density outside the bubble to be the bulk density, $\rhob$. 
The free energetics of density fluctuations (in the absence of an external field) are then given by the work that must be performed to grow the bubble against the surface tension $\gamma$ and the external pressure $\Delta \mathcal{P} = \mathcal{P} - \mathcal{P}^{\rm sat}$,
where $\mathcal{P}$ and $\mathcal{P}^{\rm sat}$ are the system pressure and the saturation pressure respectively. Thus,
\begin{align}
-\kT \ln P_v(N) &=G_0\para{r_b(N)} \nonumber \\ 
&\approx \frac{4\pi}{3}r_b^3(N) \Delta \mathcal{P} + 4\pi r_b^2(N) \gamma.
\end{align}
For the sake of simplicity, the term corresponding to the translational entropy of the bubble has been omitted from the above expression.
The radius of the bubble $r_b$ is related to $N$ by
\begin{equation}
r_b(N) = \brac{\frac{3}{4\pi\rhob}\para{\avg{N_v(\Rbar)}_0 - N}}^{1/3}.
\label{eq:rad}
\end{equation}
In the presence of attractions in $v$, work must also be done against the attractive potential,
such that
\begin{align}
-\ln P_v^{(1)}(N) &=-\ln P_v(N) + \frac{4\pi}{3} r_b^3(N) \rhob \lambda_1\beta\epsilon \\ 
&=\frac{4\pi}{3} r_b^3(N) \beta\Delta\tilde{\mathcal P} + 4\pi r_b^2(N) \beta\gamma,
\end{align}
where we have defined 
\begin{equation}
\Delta\tilde{\mathcal P}\equiv\Delta \mathcal{P}+\lambda_1\epsilon\rhob
\label{eq:ptilde}
\end{equation}
as an effective pressure in the probe volume, to illustrate that an attractive potential that couples to $\NC$ affects $P_v(N)$ in the same manner as the pressure.
Higher $\Delta \tilde{\mathcal P}$ results in an increase in magnitude of the slope at low $N$,
consistent with the results in Figure~\ref{fig:densflucs}a.
Conversely, repulsive potentials, corresponding to negative values of $\lambda_1$,
should decrease $\Delta\tilde{\mathcal P}$ and make it easier to create a cavity in the liquid.
%


\bibliography{MakeArxivMerged}

\begin{thebibliography}{85}
\bibitem [{fno()}]{fnote_typ-attr}%
  \BibitemOpen
  \href@noop {} {}\bibinfo {howpublished} {We estimate an upper bound for the
  magnitude of solute-water attractions to be
  $\lambda_i\beta\epsilon\approx0.7$. By comparison, a square-well potential
  for the $R_{\rm C}=0.6$~nm solute, with a well-depth,
  $\lambda_i\beta\epsilon\approx0.71$, has the same second virial co-efficient
  as the highly attractive C$_{60}$-water interaction potential.}\BibitemShut
  {Stop}%
\end{thebibliography}

\clearpage
\newpage

\setcounter{figure}{0}

\title{Supplementary Material for ``Water Density Fluctuations Relevant to Hydrophobic Hydration are Unaltered by Attractions''}

\author{Richard C. Remsing}
\author{Amish J. Patel}
\email{amish.patel@seas.upenn.edu}
\affiliation{Department of Chemical and Biomolecular Engineering,
University of Pennsylvania, Philadelphia, PA 19104}

\date{\today}
\maketitle


\section{Derivation of Equations 5 and 6 of the Main Text}
The probability distribution of the number of waters in the core and shell volumes can be written as
\begin{align}
P&_{v,v_{\rm sh}}^{(i)}(N,N_{\rm sh}) = \avg{\delta_{N,\NC}\delta_{N_{\rm sh},\NS}}_{\lambda_i} \\
& = \frac{\mathcal{Z}_0}{\mathcal{Z}_i} \avg{\delta_{N,\NC}\delta_{N_{\rm sh},\NS} e^{-\beta\lambda_i U(\NC,\NS)}}_0,
\end{align}
where $\mathcal{Z}_0$ and $\mathcal{Z}_i$ are the partition functions in the absence and presence of the attractive potential, respectively.
The delta functions allow the operators $\NC$ and $\NS$ to be replaced by the corresponding variables $N$ and $N_{\rm sh}$,
so that
\begin{align}
P_{v,v_{\rm sh}}^{(i)}(N,N_{\rm sh}) &= C\avg{\delta_{N,\NC}\delta_{N_{\rm sh},\NS}}_0 e^{-\beta U(N,N_{\rm sh})} \\
&=C\PNCNS e^{-\beta U(N,N_{\rm sh})},
\end{align}
where $C\equiv \mathcal{Z}_0/\mathcal{Z}_i$.
Equation 5 of the main text is then obtained by summation over $N_{\rm sh}$.
The free energy $\Delta G_i$ of turning on an attractive potential in bulk water is
\begin{align}
e^{-\beta\Delta G_i} = \frac{\mathcal{Z}_i}{\mathcal{Z}_0} = \avg{e^{-\beta \lambda_i U(\NC,\NS)}}_0.
\end{align}
By inserting the identity twice, we obtain
\begin{align}
e^{-\beta\Delta G_i}&=\bigg< \sum_N \sum_{N_{\rm sh}} \delta_{N,N_v}\delta_{N_{\rm sh},N_{v_{\rm sh}}} e^{-\beta \lambda_i U(N_v,N_{v_{\rm sh}})} \bigg>_0 \\
&=\sum_N \sum_{N_{\rm sh}} \avg{ \delta_{N,N_v}\delta_{N_{\rm sh},N_{v_{\rm sh}}} }_0 e^{-\beta \lambda_i U(N,N_{\rm sh})} \\
&=\sum_N \sum_{N_{\rm sh}} \PNCNS e^{-\beta \lambda_i U(N,N_{\rm sh})},
\end{align}
resulting in Equation 6 of the main text.


\section{Density fluctuations in the presence of attractions in small and large volumes}

\begin{figure}[bh]
\begin{center}
\includegraphics[width=0.48\textwidth]{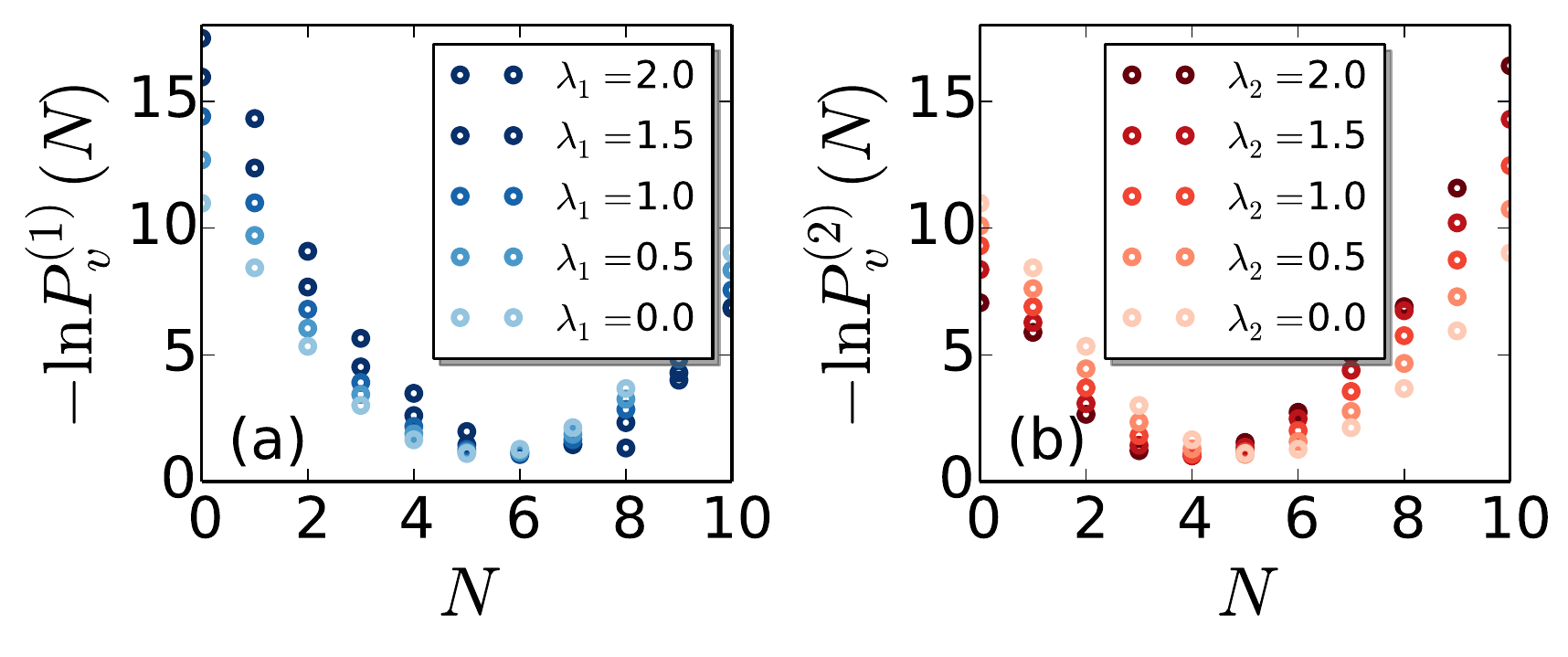}
\end{center}
\vspace{-0.5cm}
\caption[Density Fluctuations in Methane Volume]
{
Density fluctuations in the presence of the attractive fields (a) $\lambda_1 U_1$ and (b) $\lambda_2 U_2$ for a spherical solute
with $R_{\rm C}=0.336$~nm and $R_{\rm S}=0.636$~nm.
}
\label{fig:mepvn}
\end{figure}

\begin{figure}[tb]
\begin{center}
\includegraphics[width=0.48\textwidth]{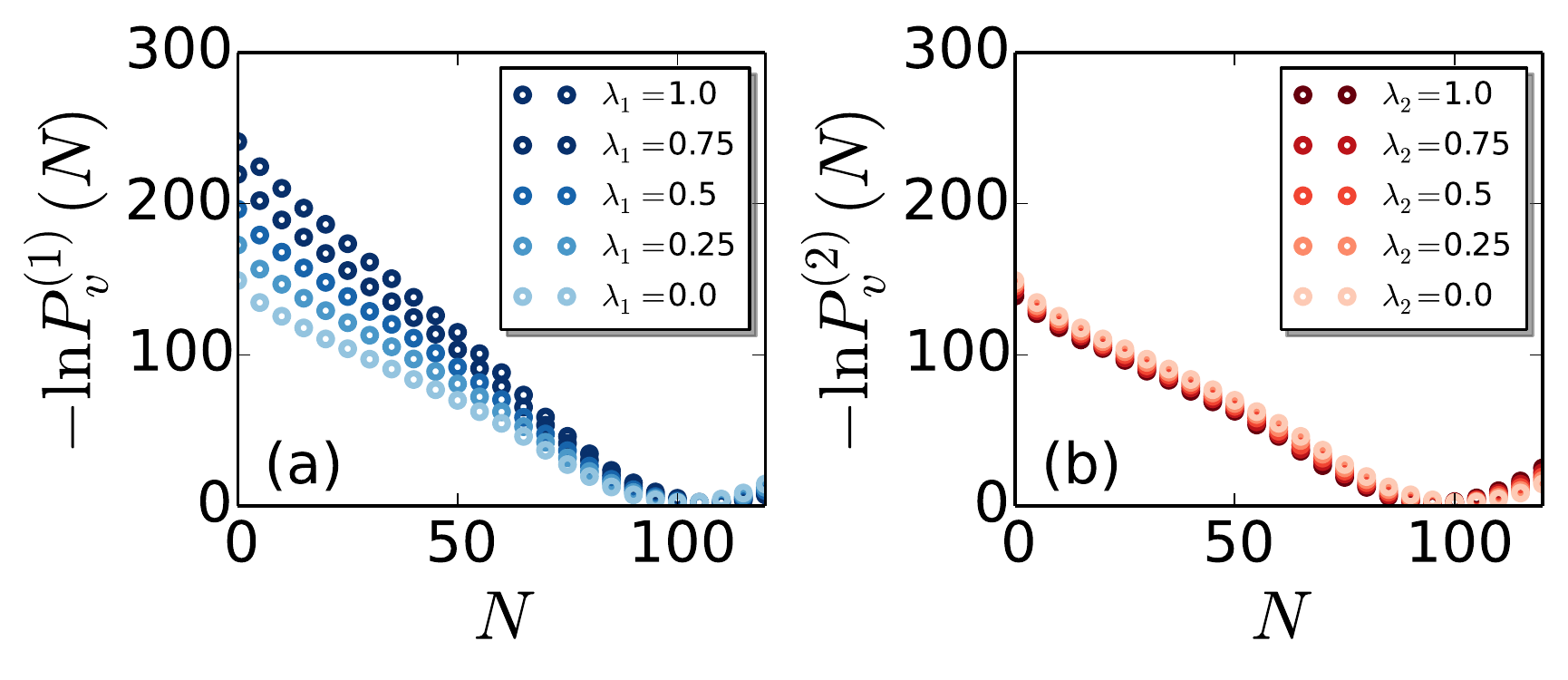}
\end{center}
\vspace{-0.5cm}
\caption[Density Fluctuations in Large Volume]
{
Density fluctuations in the presence of the attractive fields (a) $\lambda_1 U_1$ and (b) $\lambda_2 U_2$ for a spherical solute
with $R_{\rm C}=0.9$~nm and $R_{\rm S}=1.2$~nm.
}
\label{fig:r12pvn}
\end{figure}

In Figure~\ref{fig:mepvn} and Figure~\ref{fig:r12pvn}, we show the distributions
$P_{v}^{(1)}(N)$ and $P_{v}^{(2)}(N)$ for
the methane sized solute and a solute with $R_{\rm C}=0.9$~nm, respectively.
In both cases, the presence of the attractive potential, $\lambda_1 U_1$, decreases the probability of the low $N$ tails of the distribution,
independent of whether the unperturbed distribution is Gaussian, as in Figure~\ref{fig:mepvn}a,
or non-Gaussian, as in Figure~\ref{fig:r12pvn}a. 
In contrast, water density fluctuations around the mean are unaltered by the presence of attractions in the hydration shell alone, $\lambda_2 U_2$, and the distributions obtained for both the methane-sized volume (Figure~\ref{fig:mepvn}b)
and the large $R_{\rm C}=0.9$~nm volume (Figure~\ref{fig:r12pvn}b) only exhibit small decreases
in $\avg{N_v(\Rbar)}_{\lambda_2}$ with increasing $\lambda_2$. 
These findings support the conclusions drawn in the main text.

\section{Lennard-Jones Attractions}
 \label{sec:lj} 
 
\begin{figure*}
\begin{center}
\includegraphics[width=0.9\textwidth]{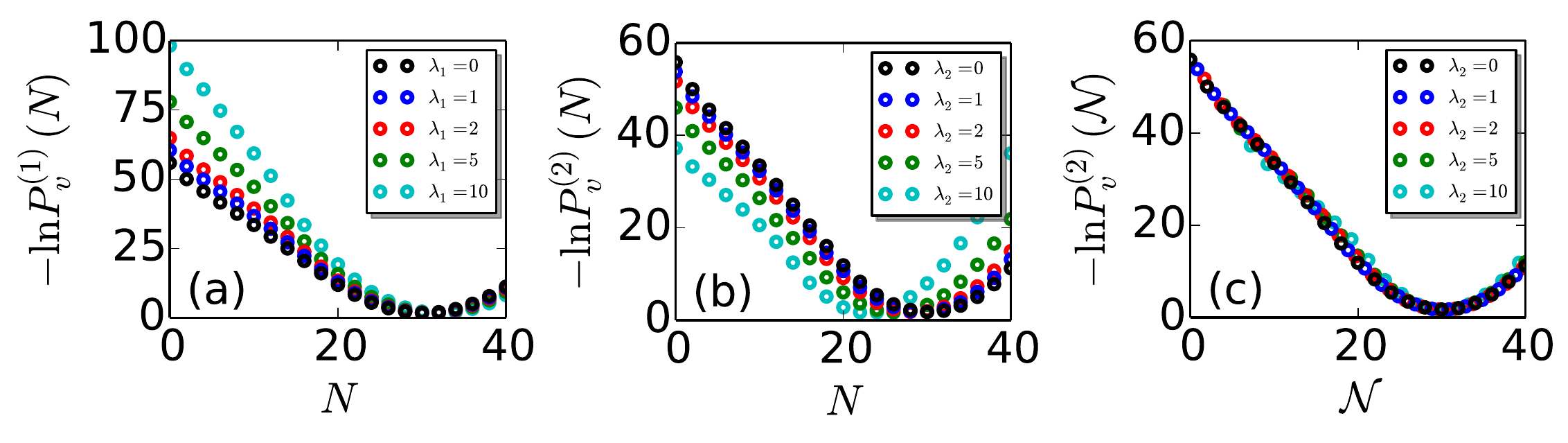}
\end{center}
\caption[Density Fluctuations with LJ Potential]
{
Water number fluctuations, $-\ln P^{(i)}_v(N)$, in the core region of a solute with radius, $R_{\rm C}=0.6$~nm, in the presence of attractive fields corresponding to the (a) full LJ attractions and (b) LJ attractions acting in the hydration shell alone.
(c) The distributions corresponding to attractions in the shell alone,
are shifted horizontally to align their means, highlighting that density fluctuations are not modified by LJ attractions in the hydration shell.
}
\label{fig:ljfluc}
\end{figure*}

 %
Here, we demonstrate that the conclusions drawn in the main text for a square-well potential are generally applicable.
Following Huang and Chandler~\cite{Huang:JPCB:2002}, we use the integrated 9--3 LJ potential, separated into purely repulsive and
 purely attractive components, $u_0^{\rm LJ}(r)$ and $u_1^{\rm LJ}(r)$, respectively, following the WCA prescription~\cite{wca}.
 We then linearly couple the attractive potential to the parameter $\lambda_1$ in order to tune the strength of the attractive
 interactions, and study how the application of the corresponding attractive potential, $\lambda_1 U_1^{\rm LJ}$, alters density fluctuations in the bulk.
The LJ-like analog of the $U_2$ potential is then obtained by applying an additional potential, $U_3$, which subtracts the attractions imposed by $U_1^{\rm LJ}$ in the solute core, as follows:
 \begin{equation}
  U_2^{\rm LJ}(\Rbar) =  U_1^{\rm LJ}(\Rbar) + U_3(\Rbar) = U_1^{\rm LJ}(\Rbar) -u_{\rm min} N_v(\Rbar),
 \end{equation}
where $u_{\rm min}$ is the minimum of the solute-solvent pair potential.
We then linearly couple this potential to a parameter $\lambda_2$ in order to vary the magnitude of the attractive interactions.
The core region, $v$, is now defined by the effective hard core radius of the solute, which is estimated as~\cite{Blip,Huang:JPCB:2002}.
\begin{equation}
R_{\rm C}\approx\int_0^\infty dr \curly{1-\exp\brac{-\beta u_{0}^{\rm LJ}(r)}}.
\label{eq:hsrad}
\end{equation}
 The probability $P^{(2)}_v(N)$ in the presence of this attractive potential, $\lambda_2 U_2^{\rm LJ}$, is then determined
 through reweighting the corresponding distribution obtained in the presence of $\lambda_1 U_1^{\rm LJ}$.
 Thus, for $\lambda_2=\lambda_1$,
 \begin{equation}
 P^{(2)}_v(N) = C P^{(1)}_v(N) e^{\beta \lambda_1 u_{\rm min}N},
 \end{equation}
 where $C$ is a normalization constant.
Density fluctuations in the absence and presence of the attractive potentials in the solute core are shown in Figures~\ref{fig:ljfluc}a and~\ref{fig:ljfluc}b, respectively, for a solute with $R_{\rm C}=0.6$~nm, and display the same qualitative trends as the square well potential.

\section{Derivation of the linear response approximation LR3 }
In order to illustrate the connection of LR3 to the other linear response approximations, we start with the exact expression for the free energy for turning on the potential, $\lambda_i U_i(\Rbar)$, in bulk water:
\begin{equation}
\beta\Delta G_i = -\ln\avg{e^{-\beta \lambda_i U_i(\Rbar)}}_0,
\label{eq:exact0}
\end{equation}
where $\avg{\cdots}_0$ is the ensemble average over configurations $\Rbar$ obtained in the absence of the attractive potential.
The linear response approximation LR2 is obtained from a second order cumulant expansion of the right hand side of Equation~\ref{eq:exact0},
\begin{equation}
\beta\Delta G_i \approx \beta\lambda_i \avg{U_i (\Rbar)}_0 -\frac{\beta^2 \lambda_i^2}{2}\avg{(\delta U_i(\Rbar))^2}_0.
\label{eq:lr2}
\end{equation}
Because the potentials considered here have the form, $U_i(\Rbar)=\epsilon N_{v_i}(\Rbar)$, the free energy becomes
\begin{equation}
\beta\Delta G_i \approx \lambda_i\beta\epsilon\avg{N_{v_i}(\Rbar)}_0 -\frac{(\lambda_i\beta\epsilon)^2}{2}\avg{(\delta N_{v_i}(\Rbar))^2}_0.
\label{eq:lr2N}
\end{equation}

We can now use the macroscopic relation between particle number fluctuations and the isothermal compressibility,
$\avg{(\delta N_{v_i}(\Rbar))^2}_0 = \rhob \kT \kappa_T \avg{N_{v_i}(\Rbar)}_0$, in conjunction with $\avg{N_{v_i}(\Rbar)}_0=\rhob v_i$,
to obtain our final expression for LR3,
\begin{equation}
\Delta G_i \approx \lambda_i\epsilon\rhob v_i-\frac{(\lambda_i\epsilon\rhob)^2}{2} \kappa_T v_i,
\label{eq:lr31}
\end{equation}
where $v_i$ is the volume over which the potential is applied.
The use of macroscopic compressibility in Equation~\ref{eq:lr31} is rigorously true only for
large (infinite) volumes.
Equation~\ref{eq:lr31} is specific to our model square well attraction potentials;
a general form applicable to all slowly-varying potentials is derived in the next section.

\section{Linear response theory for turning on slowly-varying potentials}
\label{sec:lrt}

The derivation of the generalized analog of LR3 follows the work of Remsing and Weeks~\cite{RemsingThesis} for charged systems.
For generality, consider turning on an attractive potential $w_1^{(\lambda)}(\rb)$ in a uniform fluid,
where $w^{(\lambda)}_1(\rb)$ is a slowly-varying attractive potential coupled to a parameter $\lambda$ in some fashion
that is not necessarily linear.
Within the linear response regime, the free energy of turning on the attractive potential is given by
\begin{equation}
\Delta G_\lambda \approx \frac{1}{2}\brac{\avg{ W^{(\lambda)}_1(\Rbar)}_0+\avg{ W^{(\lambda)}_1(\Rbar)}_\lambda},
\label{eq:gaussu1}
\end{equation}
where $W^{(\lambda)}_1(\Rbar)=\sum_{i=1}^{N_{\rm tot}} w^{(\lambda)}_1(\rb_i;\Rbar)$ is the total energy due to the interaction of the $N_{\rm tot}$ particles
in the system with the attractive potential. 
Such an approximation has been shown to be exceptionally accurate when
$w^{(\lambda)}_1(\rb)$ is \emph{slowly-varying} over typical correlation lengths in the fluid~\cite{RemsingThesis}.

$\Delta G_\lambda$ can be equivalently written in terms of the non-uniform singlet density,
\begin{equation}
\Delta G_\lambda = \frac{1}{2}\int d\rb \brac{\rho_{\lambda}(\rb)+\rhob} w^{(\lambda)}_1(\rb),
\end{equation}
where $\rho_\lambda(\rb)=\avg{\rho(\rb;\Rbar)}_\lambda=\avg{\sum_{i=1}^{N_{\rm tot}} \delta (\rb-\rb_i(\Rbar))}_\lambda$ is the
average non-uniform density in the presence of the field $w^{(\lambda)}_1(\rb)$, and $\rhob$ is bulk density.
To approximate $\rho_\lambda(\rb)$, we perform a functional expansion of the density, such that the deviation from the bulk density, $\Delta\rho_\lambda(\rb)\equiv\rho_\lambda(\rb)-\rhob$ is
\begin{equation}
\Delta\rho_\lambda(\rb)\approx -\beta \int d\rb' \chi_0(\len{\rb-\rb'}) w^{(\lambda)}_1(\rb),
\end{equation}
which, in Fourier space, is equivalent to
\begin{equation}
\Delta\hat{\rho}_\lambda(\kb)\approx -\beta\hat{\chi}_0(\kb)\hat{w}^{(\lambda)}_1(\kb).
\end{equation}
Here, $\chi_0(\rb,\rb')=\avg{\delta\rho_0(\rb;\Rbar)\delta\rho_0(\rb';\Rbar)}_0$ is the bulk density-density correlation function, and
$\delta\rho_0(\rb;\Rbar)=\rho(\rb;\Rbar)-\rhob$.
Because of the slowly-varying nature of $w^{(\lambda)}_1(\rb)$,
we expect only small values of $k$ to contribute to the density response and therefore the free energy.
In the limit of small $k$, the linear response function is
\begin{equation}
\hat{\chi}_0(\kb)\sim \beta^{-1}\rhob^2 \kappa_T,
\end{equation}
where $\kappa_T$ is the isothermal compressibility of the fluid.
The density response can then be approximated by
\begin{equation}
\Delta \hat{\rho}_\lambda(\kb) \approx -\beta \hat{\chi}_0^{(0)} \hat{w}^{(\lambda)}_1(\kb) = -\rhob^2\kappa_T \hat{w}^{(\lambda)}_1(\kb),
\end{equation}
to zeroth order in $k$.
Thus, in real space, the induced density response is
\begin{equation}
\rho_\lambda(\rb)\approx \rhob - \rhob^2\kappa_T w^{(\lambda)}_1(\rb).
\end{equation}
We can now use this approximation for the density response to determine the free energy, yielding
\begin{equation}
\Delta G_\lambda \approx -2a_\lambda \rhob - \frac{ \rhob^2 \kappa_T}{2} \int d\rb \para{w^{(\lambda)}_1(\rb)}^2,
\label{eq:lr3}
\end{equation}
where
\begin{equation}
a_\lambda\equiv-\frac{1}{2}\int d\rb w^{(\lambda)}_1(\rb),
\end{equation}
in analogy with the van der Waals constant.
%
%

\begin{figure}
\begin{center}
\includegraphics[width=0.45\textwidth]{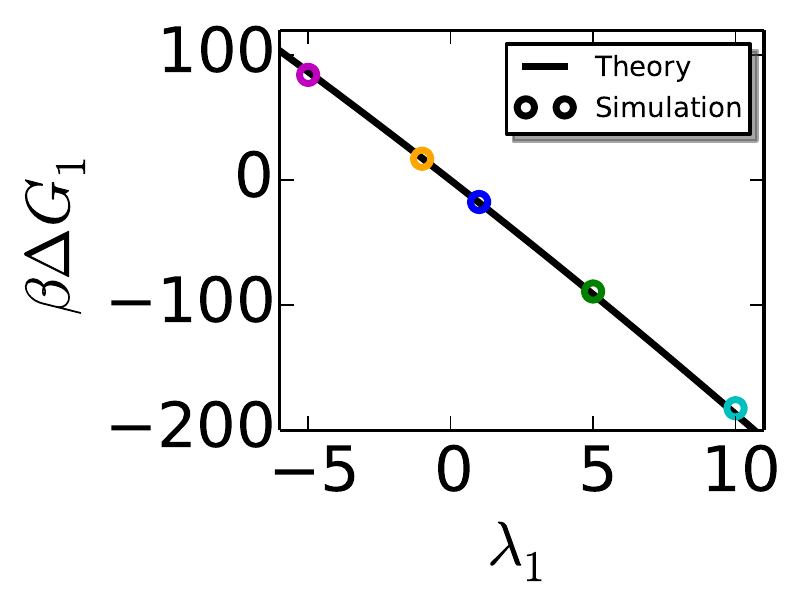}
\end{center}
\caption[Density Fluctuations with LJ Potential]
{
The free energy of turning on the 9-3 LJ attractive potential $\lambda_1 u_1^{\rm LJ}(r)$ can be accurately described by the
linear response theory Equation~\ref{eq:lr3}, as shown here for the $R_{\rm C}=0.6$~nm spherical solute.
}
\label{fig:ljtheory}
\end{figure}

%
We test these linear response theory predictions for the free energy of turning on the integrated LJ attractive potential $u_1^{\rm LJ}(r)$
in the bulk solvent.
Specifically, Figure~\ref{fig:ljtheory} shows the free energy of turning on $\lambda_1 u_1^{\rm LJ}(r)$ for a solute
with radius $R_{\rm C}=0.6$~nm, both from simulation and from Equation~\ref{eq:lr3}.
The theoretical predictions are in very good agreement with simulation results; the slowly-varying nature of the LJ potential makes it even more amenable to treatment with linear response theory than the square-well potential studied in the main text.
Additionally, because water is highly incompressible (small $\kappa_T$), we find that for typical values of $\lambda_1$, the term linear in $\lambda_1$ in Equation~\ref{eq:lr3} dominates.
%


\end{document}